
\input harvmac
\def\inbar{\,\vrule height1.5ex width.4pt depth0pt}
\def\IB{\relax{\rm I\kern-.18em B}}
\def\IC{\relax\hbox{$\inbar\kern-.3em{\rm C}$}}
\def\ID{\relax{\rm I\kern-.18em D}}
\def\IE{\relax{\rm I\kern-.18em E}}
\def\IF{\relax{\rm I\kern-.18em F}}
\def\IG{\relax\hbox{$\inbar\kern-.3em{\rm G}$}}
\def\IH{\relax{\rm I\kern-.18em H}}
\def\II{\relax{\rm I\kern-.18em I}}
\def\IK{\relax{\rm I\kern-.18em K}}
\def\IL{\relax{\rm I\kern-.18em L}}
\def\IM{\relax{\rm I\kern-.18em M}}
\def\IN{\relax{\rm I\kern-.18em N}}
\def\IO{\relax\hbox{$\inbar\kern-.3em{\rm O}$}}
\def\IP{\relax{\rm I\kern-.18em P}}
\def\IQ{\relax\hbox{$\inbar\kern-.3em{\rm Q}$}}
\def\IR{\relax{\rm I\kern-.18em R}}
\font\cmss=cmss10 \font\cmsss=cmss10 at 7pt
\def\IZ{\relax\ifmmode\mathchoice
{\hbox{\cmss Z\kern-.4em Z}}{\hbox{\cmss Z\kern-.4em Z}}
{\lower.9pt\hbox{\cmsss Z\kern-.4em Z}}
{\lower1.2pt\hbox{\cmsss Z\kern-.4em Z}}\else{\cmss Z\kern-.4em Z}\fi}
\def\IGa{\relax\hbox{${\rm I}\kern-.18em\Gamma$}}
\def\IPi{\relax\hbox{${\rm I}\kern-.18em\Pi$}}
\def\ITh{\relax\hbox{$\inbar\kern-.3em\Theta$}}
\def\IOm{\relax\hbox{$\inbar\kern-3.00pt\Omega$}}

\font\cmss=cmss10
\def\one{\hbox{\cmss 1}}

\Title{HUTP-92/A066; UICHEP-TH/92-20}{Rings and Balls}
\bigskip
\centerline{Lee Brekke$^a$}
\vskip .2 in
\centerline{Shane J. Hughes$^b$}
\centerline{and}
\centerline{Tom D. Imbo$^a$}

\footnote{}{$^a$Department of Physics, University of Illinois at Chicago,
Chicago, IL \ 60607-7059}

\footnote{}{$^b$Lyman Laboratory of Physics, Harvard University, Cambridge, MA
02138}

\vskip .9 in

We examine the various linkings in space-time of ``ball-like'' and
``ring-like'' topological solitons in
certain nonlinear sigma models in 2+1 and 3+1 dimensions. By going to
theories where soliton overlaps are forbidden, these linkings become
homotopically nontrivial and can be studied by investigating the
topology of the corresponding configuration spaces. Our analysis
reveals the existence of topological terms which give the linking number
of the world-tubes of
distinct species of ball solitons in 2+1 dimensions, or which in 3+1 dimensions
count the number of times a ball or ring soliton threads
through the center of a ring of a different species.
We explicitly construct these terms for our models, and generalize them to
cases where soliton overlaps are no longer strictly forbidden so
the terms are no longer purely topological. One of the (3+1)-dimensional
theories we consider also has topological solitons which consist of two
rings (of distinct species) linked in space.
\Date{}
\newsec{Introduction}
The existence of linking phenomena in $3$ dimensions is
familiar from everyday experience. The importance of linking (or
``braiding'') for the quantum mechanics of point particles in 2+1 dimensions
has also been well explored. For instance, the linking of the world lines of
identical particles leads to the
possibility of fractional statistics~\ref\any{R.~Mackenzie and F.~Wilczek,
Int. J. Mod. Phys. {\bf A3} (1988) 2827\semi R.~Iengo and K.~Lechner, Phys.
Rep. {\bf 213} (1992) 179\semi and references therein.}, while the linking of
particles of {\it distinct} species leads to possible
additional phases in the quantum theory which can give rise to interesting
statistical behavior for composite particles~\ref\us{L.~Brekke, A.~F.~Falk,
S.~J.~Hughes and T.~D.~Imbo, Phys. Lett. {\bf B271} (1991) 73\semi F.~Wilczek,
Phys. Rev. Lett. {\bf 69} (1992) 132.}. In
3+1 dimensions linking of the world lines of point particles is
no longer possible, but there are new types of linking if one adds fundamental
string loops to the model~\ref\har{J.~A.~Harvey and J.~Liu, Phys. Lett.
{\bf B240} (1990) 369.}. In particular, loops can be linked with each other
in space, and point particles or loops can thread through other string loops
as a function of time. These latter time-dependent processes allow the
introduction of additional phases in the quantum mechanics of these objects.

In this paper, we look for analogous linking behavior in certain nonlinear
sigma models by studying the topology of the corresponding
configuration spaces. The role of the point particles in the quantum
mechanical case is played by topological solitons possessing
a ``ball-like'' structure. The particular
examples we explore use the $O(3)$-invariant sigma model (with target space
$S^2$) in 2+1 dimensions~\ref\ot{F.~Wilczek
and A.~Zee, Phys. Rev. Lett. {\bf 51} (1983) 2250.}\ref\cp{A.~M.~Din and
W.~J.~Zakrzewski, Phys. Lett. {\bf B146} (1984) 341\semi Y.-S.~Wu and A.~Zee,
Phys. Lett. {\bf B147} (1984) 325\semi M.~Bergeron, G.~W.~Semenoff and
R.~R.~Douglas, Int. J. Mod. Phys. {\bf A7} (1992) 2417, and references
therein.} and the $O(4)$-invariant model (or ``2-flavor Skyrme model''
with target space $S^3$) in 3+1 dimensions~\ref\sky{I.~Zahed and
G.~E.~Brown, Phys. Rep. {\bf 142} (1986) 1, and references therein.}. Both of
these models have ball-like topological solitons. In 3+1 dimensions the
$O(3)$-invariant sigma model
has topological solitons that are expected to have a ``ring-like''
structure~\ref\wuzee{H.~J.~de~Vega, Phys. Rev. {\bf D18} (1978) 2945\semi
Y.-S.~Wu and A.~Zee, Nucl. Phys. {\bf B324} (1989) 623.}, so they can play the
role of the string loops in quantum mechanics.
By taking cross products of these target spaces one can get theories
with both ring and ball solitons, or with more than one species of
ring and/or ball solitons.\foot{In all of the above models, in order to obtain
nonsingular solutions to the equations of motion (whether of ring or ball
type), stabilizing terms must be added to the lagrangians.}

One can now examine the linking of rings and balls in these field
theories. Any such linking, however, is topologically trivial
because the solitons in these
models can pass through each other (since the target space is a
cross product). This problem can be solved by changing the form of the target
space from
$A\times B$ to $A\vee B$ --- the {\it wedge} (or one-point
union) of $A$ and $B$, that is, $A$ and $B$ joined at a single
point~\ref\gw{G.~W.~Whitehead, {\it Elements of Homotopy Theory}
(Springer-Verlag, New York, 1978).}. This
may be realized by starting with an $A\times B$ model in which
spatial infinity gets mapped to the point $(a_0,b_0)$ in the target space.
Then, introduce
a potential that gives an infinite amount of energy to any
configuration {\it not} of the form $(a_0,b(\vec{x}))$ or $(a(\vec{x}),b_0)$.
In such wedge models, solitons associated with the space $A$ cannot pass
through those associated with $B$
so that a study of the topology of configuration space should
detect any linking between them. (More generally, there can be no
overlap between $A$ and $B$ energy densities.) Furthermore, there should exist
topological terms which can be added to the action of these models and
which supply the possible phases seen in the quantum
mechanical case. We explicitly construct these terms in specific models,
and also their
(nontopological) generalizations to the corresponding cross product models.

The paper is organized as follows.
In section two we review  the $O(3)$-invariant sigma model in both 2+1 and
3+1 dimensions, using the $CP^{1}$ formulation. We recall that the
(2+1)-dimensional solitons may be anyonic and that such
quantizations can be implemented by using a Hopf term. We also discuss the
ring-like structure of the solitons in 3+1 dimensions.
In section three, we consider the (3+1)-dimensional sigma models with target
spaces $S^2 \times S^3$ and $S^2 \vee S^3$, which contain
both ball and ring solitons.
We examine the time-dependent configurations where a ball soliton passes
through the center of a ring and
construct a term that counts the number of such linkings.
In section four, we examine the $S^2 \vee S^2$
sigma model in both 2+1 and 3+1 dimensions. In 2+1 dimensions, we
find noncontractible loops in configuration space that correspond to
linking the ``world-tubes'' of the two distinct species of ball solitons that
exist. In 3+1 dimensions we find a rich structure of ring solitons
and examine the various homotopically distinct loops in configuration space.
In both cases we construct topological terms
that count the number of relevant linkings. We also extend these
terms to the $S^2 \times S^2$ model.

\newsec{The $S^2$ Model in 2+1 and 3+1 Dimensions}
In the (2+1)-dimensional $O(3)$-invariant nonlinear sigma
model, the scalar field $n^{a}$, $a=1,2,3$, is constrained to lie on a sphere
$S^2 \ (n^an^a=1).$ Thus, at any fixed time
the system is described by a map from $\IR^2$ to $S^2$. To ensure
that configurations have finite energy, all points at spatial infinity must be
mapped to the same point on $S^2$. So the configuration space $Q$ of the system
is equal to $Map_* (S^2 , S^2)$ --- the set of
base point preserving maps from compactified space $S^2$ to the target space
$S^2$. Since $\pi_{0}(Q)=\pi_2(S^2)=\IZ$, this model has topological solitons
and we may write $Q=\cup_{N=-\infty}^{\infty} Q_{(N)}$, where the $N$-soliton
sector $Q_{(N)}$ consists of the maps of degree $N$ from $S^2$ to $S^2$.
The degree $N$ of a configuration may be calculated from the conserved
topological current
\eqn\one{j^\mu={1 \over 8\pi}\epsilon^{\mu\nu\lambda}\epsilon^{abc}
        n^a \partial_\nu n^b \partial_\lambda n^c,}
through $N=\int d^2 x j^0$.
A convenient description of these solitons
is afforded by the so-called $CP^{1}$ formulation of the model where the fact
that $S^2 = SU(2)/U(1)$ is used to describe the theory in terms of
an $SU(2)$-valued field together with a $U(1)$ gauge symmetry which eliminates
the extra degrees of freedom~\cp .
More explicitly, $SU(2)$ may be parametrized by
$\zeta=\pmatrix{z_1 \cr z_2} \in \IC^2$ with $|z_1|^2+|z_2|^2=1$ via
\eqn\aa{U= \pmatrix{z_1&-\bar z_2 \cr z_2&\bar z_1}.}
The $S^2$ field $n^a(x)$ corresponding to $\zeta (x)$ is
\eqn\bb{n^a = \zeta^{\dag} \sigma^a \zeta,}
where $\sigma^a$ are the Pauli matrices.
Under the local $U(1)$ symmetry in the model, $\zeta$ is multiplied by a
phase, $\zeta \to e^{i\chi (x)}\zeta$, while the $n^a$ remain fixed.
The corresponding gauge field has no kinetic term in the Lagrangian --
it is an auxiliary field which can be written as
$A_\mu = i\zeta^{\dagger}\partial_{\mu}\zeta$.
The topological current in this language is
\eqn\curr{j^\mu = {1 \over 4\pi}\epsilon^{\mu\nu\lambda}F_{\nu\lambda},} where
$F_{\nu\lambda}$ is the $U(1)$ field strength. Thus,
\eqn\solnum{N=\int d^{2}x j^{0}={1 \over 2\pi}\int_{C}\vec{A}\cdot\vec{dl},}
where $C$ is the contour at spatial infinity. Note that an $N$-soliton
configuration looks like a $U(1)$ vortex with $N$ units of ``magnetic flux''.
This vortex interpretation of the solitons is also suggested by
the long exact homotopy sequence for the Hopf fibration
$$\eqalign{U(1)\hookrightarrow  S&U(2)\cr
& \downarrow \cr & S^2,\cr}$$ which gives
\eqn\a{\pi_2(S^2) = \pi_1(U(1))=\IZ.}
This shows that the topological stability of the solitons has its origins in
nontrivial windings in the $U(1)$ sector of the theory.

Some physics also resides in the fact that the fundamental group of the zero
soliton sector $Q_{(0)}$ is nontrivial:\foot{One can also show that
$\pi_1(Q_{(N)}) = \IZ$ for any $N$. The following
interpretations of the generator also hold for any $N$.
Analogous statements apply
to the various configuration spaces considered in this paper.}
\eqn\b{\pi_1(Q_{(0)})=\pi_{3}(S^{2})=\pi_3(SU(2))= \IZ.}
This group is generated by (the homotopy class of) a
time-dependent configuration
where, out of the vacuum, a soliton of degree one and its corresponding
anti-soliton are created, the soliton undergoes a $2\pi$ rotation and then the
soliton anti-soliton pair mutually annihilate.\foot{Throughout this paper,
all rotations will be taken as counterclockwise.} (More generally, if
the above $2\pi$ rotation is assigned the integer 1 in $\pi_1(Q_{(0)})$, then a
rotation by $2\pi n$ of an $N$-soliton corresponds to
$nN^2\in\pi_1(Q_{(0)})$.) In the quantum theory,
homotopically nontrivial processes such as this may be weighted by distinct
phases, where the only requirement on these phases is that they form a
representation of $\pi_1(Q_{(0)})$. The one-dimensional
unitary representations of
$\IZ$ are labelled by an arbitrary angle $\theta$.
They are given by $m:\to e^{im\theta}$ for any $m\in\IZ$.
Quantizing using such a representation, we see that the above $2\pi$
rotation picks up the phase $e^{i\theta}$ showing that the solitons
may possess fractional spin. The above generator of $\pi_1(Q_{(0)})$ may also
be represented by the time-dependent configuration where two soliton
anti-soliton pairs of unit degree are created out of the vacuum, the
two solitons interchange position (counterclockwise), and the ``new'' soliton
anti-soliton pairs
annihilate. Hence this process also picks up the phase $e^{i\theta}$ in the
above quantum theory and we see that solitons exhibiting fractional spin also
exhibit the corresponding fractional statistics in keeping with the
spin-statistics relation~\ot .

To realize a quantization with ``statistical
angle'' $\theta$, the Hopf term $H$ may be added to the action $S$~\ot\cp ,
\eqn\cc{S \to S+\theta H.}
This term can be written as
\eqn\two{H={1\over 2\pi} \int d^3 x A_{\mu}j^{\mu} = {1\over 8\pi^2} \int d^3 x
\epsilon^{\mu\nu\lambda}A_\mu F_{\nu\lambda}.}
Evaluated on a time-dependent configuration in which
an $N$-soliton is rotated by $2\pi n$, $H$ is equal to $nN^2$.

A construction similar in spirit to the one above may be used to show that the
$O(3)$-invariant nonlinear sigma model in 3+1 dimensions also has solitons,
but of a different nature than their (2+1)-dimensional counterparts. Now the
configuration space of the theory is $Q= Map_* (S^{3} , S^{2})$ and the
soliton's existence is revealed by
\eqn\c{\pi_{0}(Q)=\pi_{3}(S^2)=\IZ.}
Again, the configuration space may be decomposed into sectors of varying
soliton number, $Q=\cup_{N=-\infty}^{\infty} Q_{(N)}$.
However, the structure of the solitons is  expected to be
that of a ring; such a homotopically nontrivial
configuration being identical to that of an $O(3)$ soliton creation, rotation
and annihilation in 2+1 dimensions, only now the role of time is played by
the ``extra'' spatial dimension. The
(3+1)-dimensional time plays no role. Thus the $N=1$
soliton in the (3+1)-dimensional theory can be viewed as a ring of energy
with one unit of $U(1)$ flux running through it and also having a nontrivial
$2\pi$-twist providing topological stability~\wuzee . A ring with $n_f$
units of flux and $n_t$ $2\pi$-twists has soliton number $n_tn_f^2$.
In other words, the Hopf term $H$ in 2+1 dimensions now becomes the topological
charge in the (3+1)-dimensional model. More precisely, $H=\int d^3xh^0$ where
\eqn\three{h^{\mu}={1\over 8\pi^2}\epsilon^{\mu\nu\rho\sigma}A_\nu
F_{\rho\sigma}} is the full conserved topological current.  Note that under the
gauge transformation $A_{\mu}\to A_{\mu}-\partial_{\mu}\chi$, the current
$h^{\mu}$ changes by the total derivative $-{1\over
8\pi^2}\partial_{\nu}\epsilon^{\mu\nu\rho\sigma}\chi F_{\rho\sigma}$. However,
the topological charge $H$ is gauge invariant.
To check for the possibility of nontrivial spin, we calculate
\eqn\d{\pi_1(Q_{(0)})=\pi_{4}(S^{2})=\IZ_2.}
The generator of this $\IZ_2$ corresponds to creating a $1$-soliton
anti-soliton pair out of the vacuum, rotating the soliton by $2\pi$ and then
annihilating the pair. (This is done in (3+1)-dimensional time.)
Thus we only have the option of quantizing the ring solitons as integer
or half-integer spin objects.

\newsec{The $S^2\times S^3$ and $S^2\vee S^3$ Models}
We now consider a (3+1)-dimensional sigma model with target space $S^2
\times S^3$.
This theory has configuration space $Q= Map_* (S^{3} , S^{2}\times S^{3})$ and
\eqn\e{\pi_0(Q)=\pi_3(S^2) \times \pi_3(S^3) = \IZ \times \IZ.}
There are two types of solitons; ring solitons from the $S^{2}$ part
of the theory and ordinary ball solitons from the $S^3$ part (as in
the 2-flavor Skyrme model)~\sky .
We decompose $Q$ into $\cup_{N=-\infty}^{\infty}
\cup_{M=-\infty}^{\infty}Q_{(N,M)}$, where $N$ counts the $S^2$ soliton number
and $M$ that of $S^3$. Thus, there are two conserved topological
currents in this model; the ``Hopf current'' $h^{\mu}$ discussed above for
the $S^2$ fields, and a new current $J^{\mu}$ for the $S^3$ fields.
Since $S^3$ is the group manifold of $SU(2)$, we may view the $S^3$ fields as
$SU(2)$ matrices $U(x)$. If we define \eqn\dd{R_{\mu}=U \partial_{\mu}
U^{\dagger},} then we have \eqn\four{J^{\mu}={1 \over 24 \pi^{2}}
\epsilon^{\mu\nu\rho\sigma} tr(R_{\nu}R_{\rho}R_{\sigma}).} We also have
\eqn\f{\pi_1(Q_{(0,0)})=\pi_4(S^2) \times \pi_4(S^3) = \IZ_2 \times \IZ_2,}
and the nontrivial generator of $\pi_4(S^3)$ is represented by the creation,
rotation and annihilation of an $S^3$ 1-soliton anti-soliton pair.
Thus we may independently quantize the ring 1-soliton and the ball
1-soliton as integer or half-integer spin objects.\foot{Surprisingly, it has
been found that the baryon number 2 solutions in the 2-flavor Skyrme
model (or equivalently, the $(0,2)$ solution in the above model with an
appropriate stabilizing term) has a ring-like structure~\ref
\bra{V.~B.~Kopeliovich and B.~E.~Stern, JETP Lett. {\bf 45} (1987) 203\semi
J.~J.~M.~Verbaarschot, T.~S.~Walhout, J.~Wambach and H.~W.~Wyld,
\hfill\break Nucl.
Phys. {\bf A468} (1987) 520\semi E.~Braaten and L.~Carson, Phys. Rev. {\bf D38}
(1988) 3525\semi M.~Atiyah and N.~S.~Manton, Cambridge University preprint
DAMTP 92-32 (1992).}. However, unlike the
ring solitons we consider, one cannot treat these objects as having a $U(1)$
flux running through them since $\pi_2(S^3)$ is trivial. In what follows,
we will treat all $S^3$ solitons as ball-like.}

In this theory, there is the interesting possibility of
sliding  a ball soliton through the center of a ring.
The linking number $\gamma$ of the
world-line of a point and the world-sheet of an infinitely thin closed string
is given by~\har
\eqn\five{\gamma={1\over4\pi^2}\int d^2\Sigma_{\mu\nu}(X)\int dY_\lambda
\epsilon^{\mu\nu\lambda\sigma}{(X-Y)_\sigma \over |X-Y|^4}.}
Here $d^2\Sigma_{\mu\nu}=d\sigma d\tau\epsilon^{\alpha\beta}
\partial_\alpha X_\mu\partial_\beta X_\nu$ is the infinitesimal area element
of the string's world-sheet, and $d Y_\lambda$ the infinitesimal element
of the point's world-line.
It is possible to add a term to the action  which, in the limit that the
ball soliton becomes a point and the ring becomes a
one-dimensional string,
reduces to this term. To see this, take $J_{\mu}$ to be the topological
current of the $S^3$ soliton and $A^{\mu}$ to be the auxiliary
gauge field of the $S^2$ ring. Then,
generalize the above area and line elements as
\eqn\ee{dY_\lambda \rightarrow J_\lambda(y) d^4 y}
and
\eqn\ff{d^{2} \Sigma_{\mu\nu}(x)  \epsilon^{\mu\nu\lambda\sigma} \rightarrow
{1 \over \pi}F^{\lambda\sigma}(x) d^4 x,}
where $F^{\lambda\sigma}$ is the field strength associated with $A^{\mu}$.
Modifying the action by the resulting term
\eqn\six{\Delta S= {\theta \over 4\pi^3}\int d^4 x \int d^4 y F^{\lambda\sigma}
J_{\lambda} {(x-y)_{\sigma}  \over |x-y|^4},}
we pick up the phase exp$(in_bn_f\gamma\theta)$ every time a
ball soliton (of topological charge $n_b$) passes $\gamma$ times through the
center of a ring (with $n_f$ units of flux),
provided that there is no overlap of the energy density of the ball with that
of the ring during the process. Note that this phase is independent
of the number of $2\pi$-twists $n_t$ in the ring.
Further, upon integrating by parts and noting that
\eqn\gg{\partial^\sigma \Biggl( {(x-y)_\sigma \over |x-y|^4} \Biggr)=-{1
\over 2}\partial^\sigma
\partial_\sigma \Biggl( {1 \over |x-y|^2} \Biggr) = -2\pi^2
\delta^{(4)}(x-y),}
we see that the extra term in the action is local,
\eqn\seven{\Delta S ={\theta \over 2\pi} \int d^4 x A^{\mu}(x) J_{\mu}(x).}
In this form, $\Delta S$ can be thought of as giving the
$(0,n_b)$ ball soliton a $U(1)$ electric charge $\theta n_b/2\pi$. The factor
exp($in_bn_f\theta$) acquired upon threading a ball through the center of a
ring can now be seen as the usual Aharanov-Bohm phase exp($iq\Phi$) which
results from bringing a charge $q=\theta n_b/2\pi$ around a flux $\Phi =2\pi
n_f$. As noted above, $\Delta S$ has its topological interpretation as
$\theta n_bn_f\gamma$ only when there is
no overlap between the ball and the ring.
However, the $S^2$ and $S^3$ energy densities {\it are} allowed to overlap
and our term alters the
equations of motion. This reveals why there is no suggestion of the existence
of this term from the topology of configuration space. However, if we consider
a model in which these overlaps are forbidden, such a term should be suggested
by topological arguments. As indicated in the introduction, this can be
accomplished by choosing the
target space to be $S^2 \vee S^3$. We now turn to a discussion of this model.

The configuration space of the $S^2 \vee S^3$ model is
$Q=Map_{*}(S^3,S^2 \vee S^3)$. Solitons analogous to the ring and ball solitons
of the $S^2 \times S^3$ model exist, as revealed by\foot{The homotopy groups of
a wedge of spheres have a very rich structure. It can be shown that
$\pi_p(S^n\vee S^m)=\pi_p(S^n)\times\pi_p(S^m)\times K$ for $n,m>1$, where the
abelian
group $K$ is trivial if $p<n+m-1$ and is isomorphic to $\IZ$ if $p=n+m-1$. If
$p>n+m-1$, then $K$ can be computed using the Hilton-Milnor Theorem~\gw .}
\eqn\g{\pi_0(Q)=\pi_3(S^2 \vee S^3) =\pi_3(S^2) \times \pi_3(S^3) = \IZ \times
\IZ,} and we can write
$Q=\cup_{N=-\infty}^{\infty}\cup_{M=-\infty}^{\infty}Q_{(N,M)}.$
Since at no time can the $S^2$ and $S^3$ energy densities overlap, we expect
any time-varying configuration where a ball soliton passes through the center
of a ring to provide a nontrivial homotopy class. Indeed the  calculation of
$\pi_1(Q_{(0,0)})$ bears this out,
\eqn\h{\pi_1(Q_{(0,0)})=\pi_4(S^2\vee S^3)=\pi_4(S^2) \times \pi_4(S^3) \times
\IZ =\IZ_2 \times \IZ_2\times \IZ.} The first two factors of $\IZ_2$ are the
familiar spin factors for the two species of solitons. The factor of $\IZ$
can be generated by a time-dependent configuration in which a $(0,1)$ soliton
anti-soliton pair and a single unit flux, zero twist ring are created out of
the
vacuum, and the $(0,1)$ ball goes through the center of the ring before the
pair of ball solitons annihilate and the ring shrinks away. Thus,
in this model the term \seven\ should rise to the status of a topological
term with no effect on the classical dynamics. In fact, in the $S^2\vee S^3$
model the integrand in $\Delta S$ is locally a total derivative because
whenever $J_{\mu}$ is nonzero, $A_{\mu}$ must be pure gauge. So as expected,
the addition of $\Delta S$ does not change the equations of motion.

\newsec{The $S^2\vee S^2$ and $S^2\times S^2$ Models}
Another interesting class of sigma models we may consider are those
with target space $S^2 \vee S^2$. In 2+1 dimensions the configuration space
is $Q=Map_*(S^2,S^2\vee S^2).$ We then have
\eqn\i{\pi_0(Q)=\pi_2(S^2\vee S^2)=\IZ \times\IZ,}
and $Q$ can be written as $\cup_{N=-\infty}^{\infty}
\cup_{M=-\infty}^{\infty}Q_{(N,M)}.$
The solitons corresponding to the $(N,0)$ and $(0,M)$ sectors of $Q$ are
the corresponding $O(3)$-invariant sigma model solitons associated with the
first and second target $S^2$
respectively. These two species of solitons are distinct, so it may be useful
to assign colors to them --- say red and black --- to denote which sphere in
the target space they arise from. There are two conserved topological
currents $j_{(R)}^{\mu}$ and $j_{(B)}^{\mu}$ of the form \curr , one for each
color. Further, we find
\eqn\j{\pi_1(Q_{(0,0)})=\pi_3(S^2\vee S^2)=\IZ\times\IZ\times\IZ .}
The processes consisting of the creation, rotation and annihilation of the
red $(1,0)$ and black $(0,1)$ solitons represent generators of two of these
factors of $\IZ$. Thus, these solitons may be quantized as anyons with
independent statistical angles. Any such choice may be realized by adding to
the action both ``red'' and ``black'' Hopf terms $H^{(R)}$ and $H^{(B)}$
(as in \two ) with appropriate coefficients. The
third factor of $\IZ$ in \j\ can be generated by the creation of $(1,0)$
and $(0,1)$ soliton-antisoliton pairs, taking (say) the red soliton once
counterclockwise around the black soliton (and no other), and then
annihilating both pairs. The $2\pi$-rotation of a composite
$(1,1)$ soliton represents the element of $\pi_1(Q_{(0,0)})$ which is the
product of the above three generators. Since each of
these generators may be represented by an arbitrary phase, we see that the
spin of this composite is not uniquely determined by that of the
constituents. Now consider adding to the action the topological term
$\theta L$ where
\eqn\k{\eqalign{L &={1\over 4\pi}\int d^3 x \big (A^{(R)}_{\mu}
j^{\mu}_{(B)} + A^{(B)}_{\mu}j_{(R)}^{\mu}\big )\cr &={1\over 16\pi^2}\int d^3x
\epsilon^{\mu\nu\sigma}\big
(A^{(R)}_{\mu}F^{(B)}_{\nu\sigma} + A^{(B)}_{\mu}F^{(R)}_{\nu\sigma}
\big ).}}
Here $A^{(R)}_{\mu}$ (respectively, $A^{(B)}_{\mu}$) is the
auxiliary gauge field for the red (respectively, black) $S^2$ and
$F^{(R)}_{\nu\sigma}$ (respectively, $F^{(B)}_{\nu\sigma}$) its
corresponding field strength. This ``mixed'' Hopf term can be thought of as
providing an $(N,0)$ (respectively, $(0,N)$) soliton with a black
(respectively, red)
electric charge $\theta N/4\pi$. Evaluated on a time-dependent configuration in
which an $(N,0)$ soliton is brought $n$ times counterclockwise around a $(0,M)$
soliton we obtain $L=nNM$. Thus, $L$ can also be viewed as
the generalization to our sigma model of the linking number $n$ of two closed
curves $C_1$ and $C_2$ in $\IR^3$, which can be written
as~\ref\cc{D.~Rolfsen, {\it Knots and Links} (Publish or Perish, Berkeley,
1992).}
\eqn\eight{n={1\over4\pi}\epsilon^{ijk}\int_{C_1} dX_i\int_{C_2} dY_j
{(X-Y)_k \over |X-Y|^3}.} If we add the term
\eqn\l{\Delta S=\theta^{(R)}H^{(R)}+\theta^{(B)}H^{(B)}+\theta L}
to the action, then for a rotation by $2\pi n$ of an $(N,M)$ soliton we have
$\Delta S=nN^2\theta^{(R)}+nM^2\theta^{(B)}+nNM\theta$. The mixed Hopf term
may also be added to the $S^2\times S^2$ model, where it can be given the same
interpretation. However, unlike here, it will no longer be locally a total
derivative. Similar behavior for two species of identical particles in
(2+1)-dimensional quantum mechanics has been found in \us.

In 3+1 dimensions the $S^2\vee S^2$ model configuration space
is $Q=Map_*(S^3,S^2\vee S^2)$ and we have
\eqn\m{\pi_0(Q)=\pi_3(S^2\vee S^2)=\IZ \times\IZ\times\IZ.}
So $Q$ can be written as $\cup_{N=-\infty}^{\infty}
\cup_{M=-\infty}^{\infty}\cup_{L=-\infty}^{\infty}Q_{(N,M,L)}.$
The solitons corresponding to the red $(N,0,0)$ and black $(0,M,0)$ sectors of
$Q$ are the usual twisted rings associated with the first and second target
$S^2$ respectively. The solitons of type $(0,0,1)$ have the
structure of two untwisted unit-flux rings, one red and one black,
which link each other once. Since the
target space is $S^2 \vee S^2$, red and black flux tubes may not overlap
and these linked configurations are topologically stable.
A configuration where a ring having $n_1$ units of red flux
and $n_t$ $2\pi$-twists has linking number $n$ with a ring having $n_2$
units of black flux and $m_t$ $2\pi$-twists, belongs to the
$(n_tn_1^2,m_tn_2^2,nn_1n_2)$-th sector of $Q$. This model has three
conserved topological currents; one Hopf current of the form \three\ for each
target $S^2$, as well as an additional ``mixed'' Hopf current
\eqn\nine{\ell^{\mu}={1\over 16\pi^2}\epsilon^{\mu\nu\sigma\rho}\big
(A^{(R)}_{\nu}F^{(B)}_{\sigma\rho} + A^{(B)}_{\nu}F^{(R)}_{\sigma\rho}
\big ).}
Evaluated on the general configuration described above containing one black
and one red ring,
the quantity $\int d^3x \ell^0$ is equal to $nn_1n_2$. This topological
charge is gauge invariant although $\ell^{\mu}$ itself changes by a total
derivative under both red and black gauge transformations.

In checking for nontrivial spin and other phases we calculate
\eqn\j{\pi_1(Q_{(0,0,0)})=\pi_4(S^2 \vee S^2)=\IZ_2 \times\IZ_2 \times \IZ_2
\times\IZ \times \IZ .}
The creation, rotation and annihilation processes for the $(1,0,0)$, $(0,1,0)$
and $(0,0,1)$ solitons represent the generators of the three factors of
$\IZ_2$,
and hence we see that we may quantize these solitons as either
integer or half-integer spin objects. One of the two factors of $\IZ$ can be
generated by the process of creating from the vacuum a $(1,0,0)$ (that is, a
twisted red) soliton anti-soliton pair and a unit flux, zero twist black
ring, pulling the black ring around the red ring and then annihilating the
pair of red solitons as well as the black ring. The other factor of $\IZ$ comes
from passing a $(0,1,0)$ black ring through the center of a unit flux red ring.
We can introduce an arbitrary phase into the quantum theory for each $\IZ$.
A term in the action which supplies one such phase is
\eqn\ten{\Delta S = {\theta \over 2\pi} \int d^4x h_{(R)}^{\mu}A^{(B)}_{\mu},}
where $h_{(R)}^{\mu}$ is the red Hopf current.
Here $\Delta S$ provides
a red $(N,0,0)$ soliton with a black electric charge $\theta N/2\pi$. This
term is locally a total derivative since $A_{\mu}^{(B)}$ must be pure gauge
whenever $h^{\mu}_{(R)}$ is not zero. $\Delta S$ is not manifestly gauge
invariant, however. Under a red gauge transformation
$A^{(R)}_{\mu}\to A^{(R)}_{\mu}-\partial_{\mu}\chi$ we have
\eqn\eleven{\Delta S \to \Delta S -{\theta \over 32\pi^3} \int d^4x
\epsilon^{\mu\nu\rho\sigma} \chi F^{(R)}_{\mu\nu}F^{(B)}_{\rho\sigma}.}
Since in the $S^2 \vee S^2$ model red and black fluxes are never
allowed to overlap, the field strengths $F^{(R)}_{\mu\nu}$ and
$F^{(B)}_{\rho\sigma}$
are never nonzero at the same space-time point. Thus $\Delta S$ is invariant
under the $U(1)$ gauge transformations associated with the red $S^2$ part of
the sigma model. Under the black gauge transformation
$A^{(B)}_{\mu} \to A^{(B)}_{\mu} - \partial_{\mu} \lambda$ we obtain
\eqn\twelve{\Delta S \to \Delta S + {\theta \over 2\pi} \int d^4x
\partial_{\mu}h_{(R)}^{\mu}\lambda.}
Since the current $h_{(R)}^{\mu}$ is conserved, we see that $\Delta S$ is
also invariant under this transformation and is indeed a well-defined
term in the $S^2 \vee S^2$ sigma model. The process where an $(N,0,0)$
soliton passes through the center of a black ring with flux $m$ (and any
number of twists) gets a
phase $e^{iNm\theta}$ from this term. Similarly we may add an
analogous term to the action with the roles of the red and the black fields
interchanged, which we may use to assign an arbitrary phase to the process
where a $(0,M,0)$ soliton goes through the center of a red ring of flux $n$.

In the previous wedge models we considered, the topological terms \seven\ and
\k\ were still well-defined in the corresponding
cross product models, only in the latter these terms were
no longer of a topological nature. We may similarly ask
about the fate of \ten\ in the (3+1)-dimensional sigma model with target
space $S^{2} \times S^{2}$. This model has configuration space
$Q=Map_*(S^3,S^2\times S^2)$ and has ring solitons
associated with each target $S^{2}$, which will also be referred to
as red and black (these and all the other relevant conventions used
in the $S^{2} \vee S^{2}$ model will be followed here).
That there are these solitons and no others is shown by
\eqn\k{\pi_{0} (Q)= \pi_3(S^2)\times\pi_3(S^2) = \IZ \times \IZ,}
and thus $Q= \cup_{N=-\infty}^{\infty} \cup_{M=-\infty}^{\infty} Q_{(N,M)}$,
where $N$ counts the number of red solitons and $M$ counts the
number of black solitons. Here, of course, the linked solitons of
the $S^{2} \vee S^{2}$ model are no longer topologically stable because
the red and black fluxes can simply pass through each other. Similarly,
the topology of the configuration space does not suggest any
topological terms that yield phases when, for instance, a red soliton threads
through the hole in a black ring. More specifically,
\eqn\l{\pi_{1} (Q_{(0,0)})= \pi_4(S^2)\times\pi_4(S^2) = \IZ_{2}\times\IZ_{2},}
where the two $\IZ_2$'s indicate that the $(1,0)$ and $(0,1)$ solitons
may be independently quantized as integer or half-integer spin objects.
It is straightforward to see that $\Delta S$ in \ten\ is not even well-defined
for the $S^2\times S^2$ model.
While invariant under black gauge transformations
for the same reasons as in the $S^{2} \vee S^{2}$ case, the fact that the red
and black fluxes are now allowed to overlap means that it is
not invariant under red gauge transformations. However, all is not
lost. We may add a term to $\Delta S$ that makes the total term
well-defined (but not topological) and supplies an arbitrary phase to
the process of sliding a twisted red ring through the center of a black ring.
Indeed write $\zeta \in \IC^{2}$, where $\zeta$ parametrizes the
red $S^{2}$, as
\eqn\zz{\zeta = \pmatrix{e^{i\omega_{1}}cos(\phi) \cr
e^{i \omega_{2}}sin(\phi)}}
and define $\omega =(\omega_1 + \omega_2)/2.$
Then under the red gauge transformation
$A^{(R)}_{\mu}\to A^{(R)}_{\mu}-\partial_{\mu}\chi$ we have
$\omega \to \omega + \chi$. Note that $\omega$ is only defined modulo
$\pi$, and is completely ambiguous if either $cos(\phi)$ or $sin(\phi)$
equals zero. However, since $F^{(R)}_{\mu\nu}$ vanishes in this latter
case and
\eqn\hh{\int d^4x\epsilon^{\mu\nu\rho\sigma}F^{(R)}_{\mu\nu}F^{(B)}_{\rho
\sigma}=0,}
we see that
\eqn\thirteen{\Delta S' = {\theta \over 2\pi} \int d^{4}x
\big ( h_{(R)}^{\mu}A^{(B)}_{\mu}+{\omega\over 32\pi^3}\epsilon^
{\mu\nu\rho\sigma}F^{(R)}_{\mu\nu}F^{(B)}_{\rho\sigma}\big )}
is well-defined. Moreover,
by \eleven , $\Delta S'$ is invariant under red gauge transformations.
It remains invariant under black gauge transformations
since only the black field strength is involved in the modification.
Also, the modification is nonzero only when the field strengths of
the red and black fluxes overlap, so we see that this term still supplies the
phase $e^{iNm\theta}$ to the process where a red $(N,0)$ ring passes
through the center of a black ring with flux $m$. A similar
term, again with red and black interchanged, supplies a phase to the
passing of a twisted black ring through the center of a red ring.

\newsec{Conclusion}
In various nonlinear sigma models, we have explicitly constructed terms which
give the linking number of the world-tubes of distinct species of ball
solitons in 2+1 dimensions or which detect the threading
of balls or loops through other loops (of distinct species) in
3+1 dimensions. We also found that the (3+1)-dimensional $S^2 \vee S^2$
model has solitons consisting of two flux rings of distinct
species linked in space. The $A \vee B$ type models which appear in our
analysis may be considered
as theories in their own right by adding additional fields to smooth out
the target space singularity at the point where $A$ and $B$ are joined
(which can be done without essentially altering the topology).
Alternatively, they may be considered as limits of cross product models as
discussed earlier. In any case, they
should be generally useful for thinking about linking effects
in field theories since they reduce certain questions about such phenomena to
questions about topology.

\bigskip

\noindent
{\bf Acknowledgements}\hfil\break

\noindent
This work was supported in part by DOE contract DE-FG02-84ER40173 (L.B. and
T.I.) and by NSF grant PHY-87-14654 (S.H.). Most of this work was done
while T.I. was a Junior Fellow in the Harvard Society of Fellows.

\listrefs
\bye